\documentclass[10.5pt, twocolumn]{article}
\usepackage[left=12mm, top=15mm, bottom=15mm, right=12mm]{geometry}
\usepackage{xcolor}
\usepackage{sectsty}
\usepackage{hyperref}
\usepackage{graphicx}
\usepackage{amsmath}
\usepackage[nomove,sort]{cite}
\usepackage{comment}
\usepackage{here}
\usepackage{afterpage}
\usepackage{xspace}
\usepackage{bm}
\usepackage{braket}
\newcommand{\Th}{$^{229}$Th\xspace}
\newcommand{\Thm}{$^{229{\mathrm{m}}}$Th\xspace}
\usepackage{chemformula}
\usepackage{siunitx}
\usepackage[utf8]{inputenc}
\usepackage{placeins}
\usepackage{dblfloatfix}
\usepackage{lineno}
\title{Laser M\"{o}ssbauer spectroscopy of $^{229}$Th}

\author{
Takahiro~Hiraki{$^1$}\footnote{Corresponding author: thiraki@okayama-u.ac.jp}, Takahiko~Masuda{$^1$}, Sayuri~Takatori{$^1$}, Fabian~Schaden{$^2$}, Michael Bartokos{$^2$}, \\ 
Kjeld~Beeks{$^{2, 3}$}, Yuta~Fukunaga{$^1$}, Andreas~Grüneis{$^2$}, Ming~Guan{$^1$}, Georgy~Kazakov{$^2$}, \\ 
Thomas~LaGrange{$^3$}, Adrian~Leitner{$^2$}, Ira~Morawetz{$^2$}, Ryoichiro~Ogake{$^1$}, Koichi~Okai{$^1$}, \\ 
Martin~Pimon{$^2$}, Martin~Pressler{$^2$}, Thomas~Riebner{$^2$}, Noboru~Sasao{$^1$}, Felix~Schneider{$^2$}, \\ 
Thorsten~Schumm{$^2$}\thanks{Corresponding author: thorsten.schumm@tuwien.ac.at}, Kotaro~Shimizu{$^1$}, Luca~Toscani~de~Col{$^2$}, Tomas~Sikorsky{$^{2, 4}$}, \\ 
Akihiro~Yoshimi{$^1$}, Koji~Yoshimura{$^1$} \\[5mm] \\
{$^1$}Research Institute for Interdisciplinary Science, Okayama University, Okayama, 700-8530, Japan\\
{$^2$}Faculty of Physics, TU Wien, 1020 Vienna, Austria\\
{$^3$}Institute of Physics, Laboratory for Ultrafast Microscopy and Electron Scattering LUMES,\\ \'{E}cole Polytechnique F\'{e}d\'{e}rale de Lausanne, Station 6, CH-1015 Lausanne, Switzerland\\
{$^4$}Department of Chemical Physics and Optics, Charles University, Prague, Czechia\\
}

\begin{document}
\date{}
\maketitle

\begin{abstract}
Mössbauer spectroscopy is widely used in biochemistry, geology, and solid-state physics to obtain structural information on materials.
Here, we extend this technique into the optical range using a vacuum ultraviolet laser to probe the low-energy nuclear transitions of thorium-229, doped in calcium fluoride crystals. We discover four distinct doping sites for the thorium ions, determine the characteristic electric field gradients emerging in the interaction with the host crystal, and identify the microscopic structure of the two dominant configurations.
Site-selective laser excitation allows to study the isomeric state lifetime and laser-induced quenching for all sites. This laser-based Mössbauer spectroscopy provides a powerful probe of the nuclear environment, yielding foundational data for designing future solid-state nuclear clocks.
%\vspace{1cm}
\end{abstract}

%TC:endignore

%%%%%%%%%%%%%%%%%%%%%%%%%%%%%%%%%
\section{Introduction}
%%%%%%%%%%%%%%%%%%%%%%%%%%%%%%%%%

In solids, nuclear energy levels are split by the interaction of the nuclear electric quadrupole moment with the electric field gradient (EFG) originating from the local chemical environment. 
Spectroscopy of this nuclear splitting, known as Mössbauer spectroscopy\cite{Moessbauer1971}, is a powerful tool to investigate the local structure surrounding a nucleus. 
In conventional Mössbauer spectroscopy, nuclear excitation energies are typically in the 10–-100 keV range. 
As no narrow-linewidth light sources are currently available at these energies, this technique relies on the resonant reabsorption of $\gamma$-rays by the target nuclei. 
The $^{229}$Th nucleus, however, presents a unique case. 
It possesses a low-lying metastable isomeric state ($^{229m}$Th) accessible via laser radiation, enabling a novel approach: laser Mössbauer spectroscopy. 
Here, a narrow-linewidth laser, tunable over a wide range, can directly probe the EFG and thus the local structure of the target nuclei. 

The nuclear transition of $^{229}$Th corresponds to a vacuum ultraviolet (VUV) wavelength of approximately 148\,nm. 
This unique nuclear transition holds significant promise for applications in high-precision frequency standards, commonly referred to as nuclear clocks\cite{Peik2003,Rellergert2010,Campbell2012,Kazakov2012}. 
Solid-state nuclear clocks, offering advantages such as high dopant density and compactness, hold potential for novel applications beyond the capabilities of current atomic clocks\cite{Beeks2021}. 

Direct laser excitation of $^{229}$Th nuclei embedded in CaF$_2$ and LiSrAlF$_6$ single crystals, as well as in $^{229}$ThF$_4$ thin films, was achieved in 2024\cite{Tiedau2024,Elwell2024,Zhang2024,Zhang2024_2}. 
Using a narrow-linewidth VUV frequency comb\cite{Zhang2022}, the excitation frequency of the $^{229}$Th nucleus was measured with kilohertz-level accuracy\cite{Zhang2024,Higgins2025}.  
Furthermore, it has been demonstrated that the isomeric state population can be quenched back to the nuclear ground state using X-rays\cite{Hiraki2024,Guan2025_2} or lasers\cite{Schaden2025,Terhune2025} to accelerate a clock interrogation cycle. 

\begin{figure*}[h]
    \centering
    \includegraphics[width=19cm]{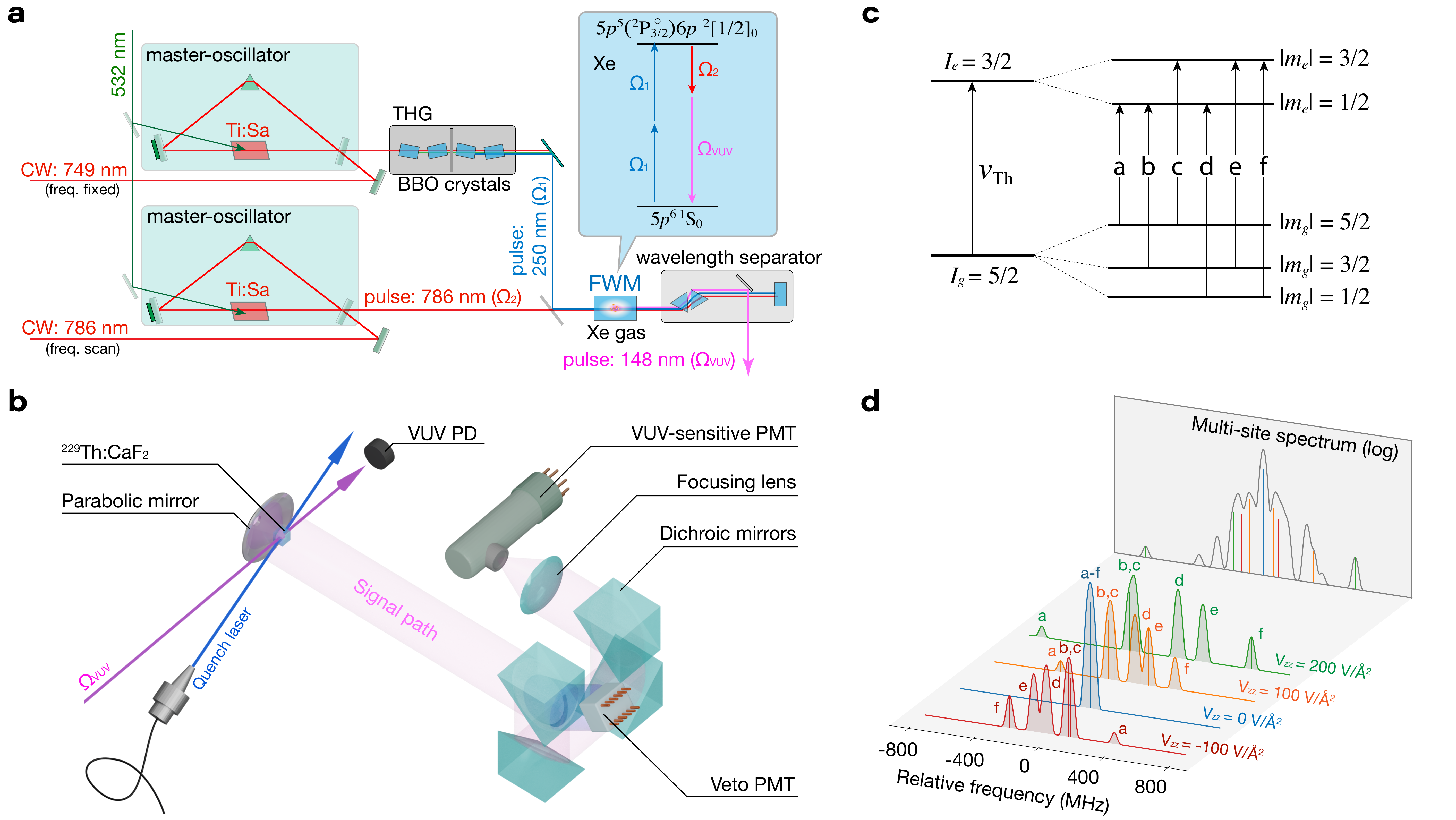}
    \caption{ {\bf Schematic view of the \Th laser spectroscopy experiment}.
\textbf{a}, The pulsed VUV laser system. 
The VUV beam diameter at the crystal target is approximately 1\,mm, comparable to the side length of the crystals. 
Motorized flippers are positioned upstream of the xenon gas chamber to block the laser from entering it during the observation of de-excitation light emitted from the crystal. 
\textbf{b}, Overview of the target and the detection system. 
A motorized rotating wheel is used to prevent scattered light from reaching the PMT while the laser irradiates the crystal. 
\textbf{c}, Nuclear energy level splitting due to the electric field gradient, assuming $V_{zz}>0$. 
\textbf{d}, Sketch of spectra when \Th is doped at more than one site. 
}
    \label{fig:setup}
\end{figure*}

The interaction with the EFG splits the nuclear energy levels. 
The $^{229}$Th ground state ($I_g = 5/2$) splits into three sublevels, while the isomeric state ($I_e = 3/2$) splits into two, resulting in six possible nuclear transitions, as depicted in Fig.~\ref{fig:setup}C. 
Conventionally, the axes of the EFG are defined so that $V_{ij}=\partial V/\partial r_i \partial r_j $ ({$r_i$} = {$x, y, z$}) is a diagonal matrix, and $\left| V_{zz} \right| \geq \left| V_{yy} \right| \geq \left| V_{xx} \right|$. 
The split frequencies are proportional to $QV_{zz}$, where $Q$ denotes the nuclear spectroscopic electric quadrupole moment. 
The level structure is further described by the asymmetry parameter $\eta = (V_{xx}-V_{yy})/V_{zz}$, which is constrained by Laplace's equation to $0\leq\eta\leq 1$. 

Currently, $^{229}$Th:CaF$_2$ is the only system in which the quadrupole splitting of $^{229}$Th has been experimentally observed\cite{Zhang2024,Higgins2025} and studied extensively\cite{Kazakov2012,Tiedau2024,Dessovic2014,Stellmer2015,Beeks2023,Beeks2024,Nalikowski2025,Kraemer2023,Pineda2025,Hiraki2024,Schaden2025,Guan2025_2}. 
Refs.\cite{Zhang2024,Higgins2025} studied one dopant site characterized by $Q_gV_{zz}=335.331(9)$\,eb\,V/$\mathring{\mathrm{A}}$$^2$ and $\eta=0.57184(5)$ at 293\,K by using direct VUV frequency comb spectroscopy. 
The ratio of the electric quadrupole moments of $^{229m}$Th to the ground state is measured to be $Q_e/Q_g=0.57003(1)$. 
The relative intensities of the transitions between these sublevels are determined by $\eta$\cite{Beeks2024_2}. 
However, the observation of unassigned transitions suggested that $^{229}$Th ions occupy multiple, previously uncharacterized sites within the host crystal lattice\cite{Zhang2024}. 

In this work, we perform laser spectroscopy of the nuclear clock transitions of $^{229}$Th doped into solid-state CaF$_2$ single crystals using a VUV pulsed laser with a full width at half maximum (FWHM) linewidth of 30 MHz. 
We employ a high signal-to-noise ratio detector system\cite{Hiraki2024,Hiraki2024_2,Guan2025}, enabling the assignment of four distinct microscopic sites, each with a characteristic EFG. 

%%%%%%%%%%%%%%%%%%%%%%%%%%%%%%%%%%
\section{Experimental apparatus}
%%%%%%%%%%%%%%%%%%%%%%%%%%%%%%%%%%
Figure~\ref{fig:setup}a provides an overview of the VUV laser setup. 
Two external cavity diode lasers (ECDLs) operating at 749\,nm and 786\,nm serve as narrow-linewidth continuous-wave (CW) seed lasers, which are injected into titanium-sapphire (Ti:Sa) ring cavities\cite{Hori2009}. 
A 10\,Hz pulsed Nd:YAG laser (Litron Nano L, 532\,nm) with pulse energies of 40--50\,mJ, pumps the Ti:Sa crystals. 
The resulting 749\,nm and 786\,nm pulses typically have output energies of 4.5--6.0\,mJ and temporal widths of 40--50\,ns (FWHM). 
The 749\,nm pulses are then injected into $\beta$-BaB$_2$O$_4$ (BBO) crystals to generate 250\,nm pulses via third-harmonic generation. 
Subsequently, by coaxially injecting the 250\,nm and 786\,nm lasers into a xenon gas cell, VUV light is generated through a four-wave mixing process resonant with a Xe transition, $5p^6\ ^1\mathrm{S}_0 \to5p^5(^2\mathrm{P}^\circ_{3/2})6p\ ^2[1/2]_0$, as previously demonstrated by refs.\cite{Thielking2023, Elwell2024}. 
The significantly narrower linewidths of the fundamental lasers used in this work compared to those in refs.\cite{Thielking2023, Elwell2024} enable us to resolve the quadrupole structures of the nuclear clock transitions. 

The VUV pulses are separated from the other wavelengths using a pair of MgF$_2$ prisms and are subsequently reflected by a D-shaped mirror on a motorized stage. 
After separation, the VUV pulse energy reaches up to 500\,nJ, although this output is drifting, primarily due to surface contamination of the MgF$_2$ optics. 
During frequency scans, the VUV intensity is monitored using a photodiode (Hamamatsu, S8552) covered by VUV band-pass filters, and is actively stabilized by adjusting the intensity of the 786\,nm pulses. 
The VUV laser frequency is tracked by monitoring the fundamental ECDLs with a wavemeter (HighFinesse, WS7). 
This wavemeter is calibrated against a 780\,nm CW laser source, locked to a rubidium D2 line.

Three crystals with different \Th concentrations are used: ``C10'' ($4\times10^{14}$ mm$^{-3}$), ``C13'' ($8\times10^{14}$ mm$^{-3}$), and ``X2'' ($5\times10^{15}$ mm$^{-3}$). 
C13 and X2 were annealed at 1250\,$^\circ$C in a CF$_4$ atmosphere to improve VUV transmittance\cite{Beeks2024}. 
Each crystal is cut into a cuboidal shape of about 1\,mm$^3$ and mounted on a holder with thin metal wires. 
The position is adjusted using a three-dimensional motorized stage. 

A custom-built signal detection system\cite{Hiraki2024,Hiraki2024_2,Guan2025} is employed, as shown in Fig.~\ref{fig:setup}b. 
In the \Th-doped crystals, $\alpha$- and $\beta$-decays of $^{229}$Th and its daughter nuclides cause bursts of photons, a phenomenon known as radioluminescence. 
The crystal position is optimized by maximizing the rate of these radioluminescence background events. 
Throughout this experiment, the crystal is maintained at room temperature. 

Light emitted from the crystal is collimated by a parabolic mirror.
Background photons with broad spectra, originating from radioluminescence~\cite{Beeks2022,Stellmer2015}, are significantly suppressed by spectral filtering.
This is achieved using four right-angle dichroic mirrors with custom coatings, designed to have a reflectance peak around 150\,nm, arranged orthogonally to each other.
The photons reflected by these mirrors are subsequently focused by an MgF$_2$ lens and detected by a solar-blind photomultiplier tube (PMT, Hamamatsu R10454). 

Radioluminescence background is further suppressed by temporal filtering using an additional PMT (Hamamatsu R11265-203) installed behind the first right-angle dichroic mirror. 
This PMT measures the timing of photon bursts, and signals detected in coincidence by both PMTs are rejected as radioluminescence.  
During laser irradiation, a motorized wheel blocks the light path to the PMTs to prevent damage. 
An oscilloscope (National Instruments, PXIe-5162) records the amplified waveforms from the PMTs, the signal from a pulse generator which is used for monitoring data acquisition and background rejection efficiencies, and the signal waveform from the photodiode. 
This setup achieves a selection efficiency for \Thm de-excitation signals exceeding 94\% in C10 and C13 and 45\% in X2, while eliminating over 97\% and 99\% of background events, respectively. 
After background rejection is applied, the approximate background rates are 0.2\,cps for C10, 0.3\,cps for C13, and 1.0\,cps for X2. 
The maximum detection rate is controlled by varying the laser irradiation period to stay below the data acquisition system's limit of approximately 1000\,Hz. 

% \clearpage
%%%%%%%%%%%%%%%%%%%%%%%%%%%%%%%%%%%
\section{Spectroscopic measurement and identification of microscopic sites}
\begin{figure}[htbp]
\centering
\includegraphics[width=9.0cm]{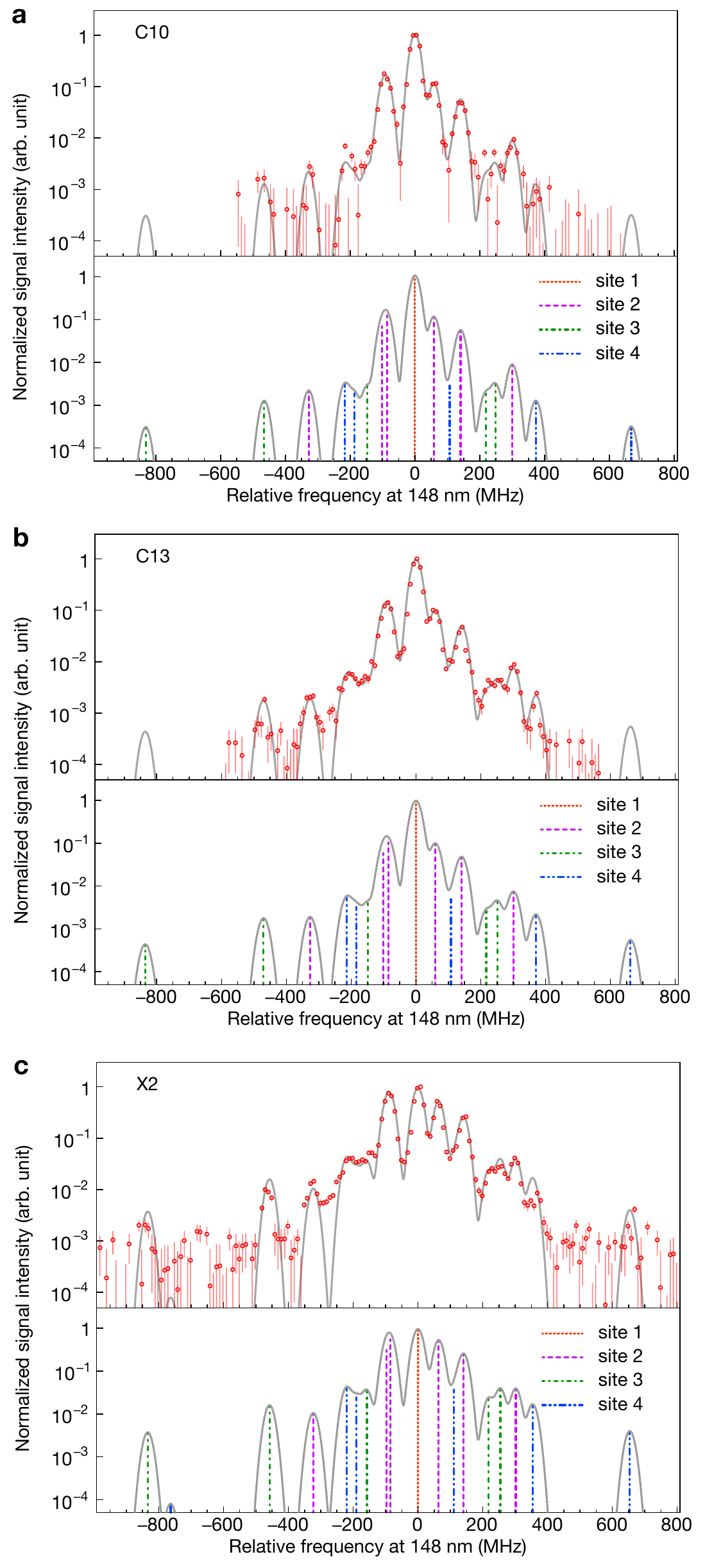}
\caption{{\bf Comparison of the measured spectra and fit results for C10 (\textbf{a}), C13 (\textbf{b}), and X2 (\textbf{c})}. 
The height of each line in the bottom panels indicates the transition intensity obtained from the fit. 
The center frequencies of site~1 peaks are overlapped, and the corresponding lines are combined into a single line representing the summed intensity. 
The error bars represent the 68\% confidence interval of the statistical uncertainty. 
}
\label{fig:spectra_fitted}
\end{figure}

%%%%%%%%%%%%%%%%%%%%%%%%%%%%%%%%%%%
The VUV laser frequency is scanned in 10\,MHz steps around the $^{229}$Th excitation frequency ($\nu_{\mathrm{Th}}\approx2020407$\,GHz) by tuning the frequency of the 786\,nm ECDL, while the frequency of the 749\,nm ECDL remains fixed. 
For C10 and C13, a frequency range of 1.2\,GHz is scanned, while for X2 a range of 1.8\,GHz is scanned. 
Each data point corresponds to a 60\,s irradiation period followed by a 300\,s detection period. 
This detection period is shorter than the radiative lifetime of \Thm, hence a residual signal remains in successive measurements, which is removed from the data in offline analysis (see Methods). 
The obtained spectra for C10, C13, and X2 are presented in the top panels of Fig.~\ref{fig:spectra_fitted}a--c. 
We observe a multitude of nuclear lines with signal amplitudes spanning three orders of magnitudes. 
The observed line widths of approximately 30\,MHz are determined by the laser linewidth. 

The observed spectra can be explained and fitted assuming that \Th is embedded in the CaF$_2$ crystal at four distinct sites, each with a characteristic EFG. 
The $V_{zz}$ are roughly 0\,V/\AA$^2$, $110$\,V/\AA$^2$, $-320$\,V/\AA$^2$, and 260\,V/\AA$^2$ for sites 1, 2, 3, and 4, respectively. 
At site 2, all six peaks are observed and the $V_{zz}$ value coincides with that reported in refs.\cite{Zhang2024, Higgins2025}. 
The spectra were fitted assuming a Gaussian lineshape for each peak and the presence of \Th at these four distinct sites. 
Fit parameters of each site~$s$ include $V_{zz}$, $\eta$, the unsplit transition frequency $u_s$, and relative \Th doping amount $a_s$. 
The peak width $\sigma$, assumed to be Gaussian, is a common parameter for all sites and lines. 
The ratio of the electric quadrupole moments $Q_e/Q_g$ is the remaining fit parameter, where $Q_g=3.11$\,eb is used as a fixed input\cite{Porsev2021}. 
The fit function for each frequency spectrum is given by 
\begin{align}
    \mathrm{func}(f)=\sum_{s=1}^4\sum_{i=1}^6 a_s r_i\exp\left(-\frac{(f-u_s-f'_{s,i})^2}{2\sigma^2}\right),
\end{align}\label{eq:fitfunc}
where $r_i = r(i,\eta)$ is the relative transition intensity of the $i$th transition and $f'_{s,i} = f'(s,i,V_{zz},\eta,Q_e/Q_g)$ is the frequency shift of the $i$th transition of each site. 
Calculation of $r_i$ and $f'_{s,i}$ are summarized in the Methods section. 
For $\eta =0$, the transition between $m = \pm 5/2\rightarrow m'=\pm 1/2$ is asymptotically forbidden. 
Figure~\ref{fig:spectra_fitted} shows a comparison between the measured spectra and the fit results. 

%%%%%%%%%%%%%%
\subsection{Quantitative analysis of nuclear quadrupole spectra}
%%%%%%%%%%%%%%

Two full spectra of the \Th nuclear quadrupole structure in CaF$_2$ are recorded for all three crystals.
The average EFG parameters calculated from the fitting are summarized in Appendix. 
The extracted $V_{zz}$ parameters are consistent across the different crystals. 
The $Q_e/Q_g$ ratios obtained from the fitting are 0.574 for C10, 0.570 for C13, and 0.563 for X2.
The EFGs for site~2 and $Q_e/Q_g$ are consistent with those precisely measured in ref.\cite{Higgins2025}. 
The relative signal contributions from \Th at the four identified sites are summarized in Table~\ref{tab:relative_Th_amount}. 
Notably, the relative contribution of \Th doped at site~1 in X2 is significantly smaller than that in C10 and C13. 

\begin{table}[htbp] 
\caption{ \textbf{Relative contribution of microscopic sites to spectroscopy signal (\%)}. Values in parentheses are differences of fitting results of the two spectra.} 
\vspace{10pt}
\centering
	\begin{tabular}{ccccc}  
		\hline
		target & site~1 & site~2 & site~3 & site~4 \\
		\hline \hline
        C10 & 72.6(2) & 26.4(5) & 0.4(3) & 0.6(1) \\  
        C13 & 73.8(6) & 24.0(7) & 1.0(1) & 1.2(1)   \\    
		X2  & 34.8(3) & 58.6(12) & 3.6(4) & 3.0(12) \\
		\hline
	\end{tabular}
	\label{tab:relative_Th_amount}
\end{table}
%%%%%%%%%%%%%
\subsection{Lifetime and quenching of \Thm in different sites}
%%%%%%%%%%%%%

\begin{figure}[htbp]
\centering
\includegraphics[width=9.0cm]{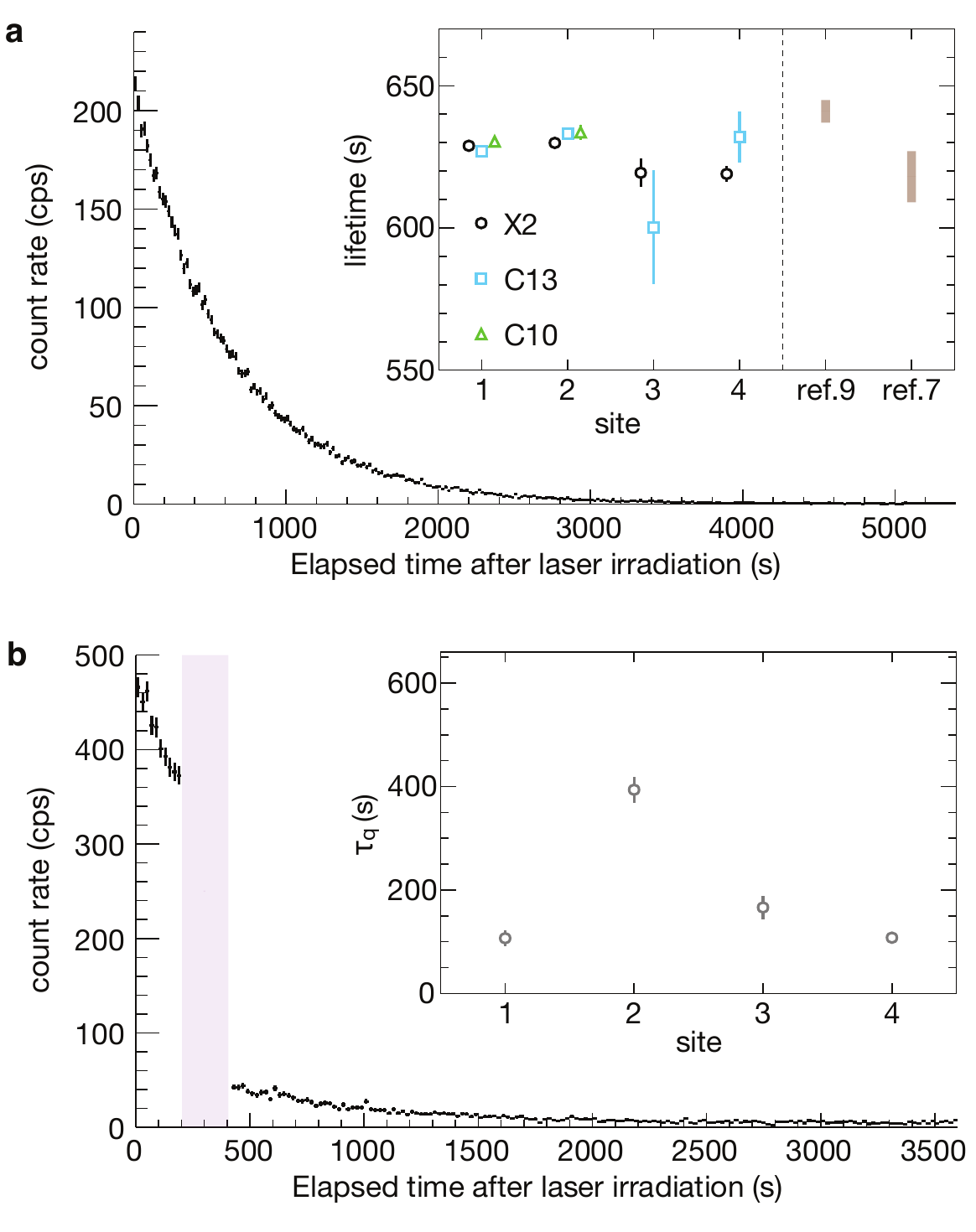}
\caption{{\bf Site-selective isomer lifetime and quenching}. \Th nuclei are excited in sites 1, 2, 3, and 4 by detuning the VUV excitation laser by 0\,MHz, $-90$\,MHz, $+240$\,MHz, and $-210$\,MHz, respectively.
\textbf{a}, Example of a radiative decay curve of \Thm in site~1 in the C13 crystal.
The inset shows a comparison of the measured lifetimes across different sites and crystals.
\textbf{b}, Example of laser-induced quenching of \Thm in site~1 in the X2 crystal. 
The shaded area indicates the time of irradiation of the crystal by the quenching laser. 
The inset shows a comparison of measured quenching lifetime $\tau_q$ for different sites of the X2 crystal. 
The error bars represent the 68\% confidence interval of the statistical uncertainty in both insets.
}
\label{fig:lifetime_summary}
\end{figure}

The direct tunability of the VUV laser allows to selectively excite \Th in the four identified crystal sites, enabling a measurement of the radiative lifetime $\tau$. 
Figure~\ref{fig:lifetime_summary}a shows a representative nuclear decay curve after background rejection, fitted with a single exponential function plus a constant offset. 
The resulting radiative lifetimes are presented in the inset of Fig.~\ref{fig:lifetime_summary}a. 
We observe lifetimes of approximately 630\,s, consistent with previously reported values\cite{Zhang2024,Schaden2025}. 
No apparent dependence of $\tau$ on the doping site was observed.

Laser-induced quenching (LIQ) is also measured using the X2 crystal. 
Any potential quenching effect from the VUV excitation laser on the frequency scan measurement is considered negligible due to its weak intensity\cite{Schaden2025}. 
For the LIQ measurement, a 405\,nm CW laser source serves as the quenching source; it is placed diagonally below the target crystal, as shown in Fig.~\ref{fig:setup}b.  
Its average power is roughly 10\,mW, although power fluctuations makes a precise estimation of the intensity on the crystal difficult. 
Figure~\ref{fig:lifetime_summary}b shows an example of a decay curve before and after the quenching laser irradiation. 
The resulting quenched lifetimes ($\tau_q$), presented in the inset, clearly show a dependence on the microscopic site. 
This is in stark contrast to $\tau$, which is site-independent. 

\subsection{Assignment of atomic structure to sites} 

The central frequency peak without discernible nuclear substructure observed in the laser Mössbauer spectroscopy implies that the microscopic site 1 has a vanishing EFG on the \Th position and therefore a high symmetry of the local chemical environment. Although the pristine CaF$_2$ host crystal lattice exhibits the necessary $O_h$ symmetry, defects such as vacancies, interstitials, or impurities generally reduce this symmetry.

Recent experimental studies have ruled out interstitial thorium and shown a 4+ charge state~\cite{Kraemer2023, Takatori2025}, leading us to focus on thorium substituting for calcium (\ch{Th_{Ca}} in Kröger-Vink notation~\cite{Kroger1956}) as the dominant defect mechanism. Through this substitution, two of thorium's four valence electrons are loosely bound and will likely be compensated by the crystal to achieve a low-energy closed-shell configuration. The primary compensation mechanisms involve either a calcium vacancy (\ch{v_{Ca}}), two fluorine interstitials (\ch{2 F_i}) or oxygen impurities (\ch{O_F}, \ch{O_i}). However, placing these compensations on nearest neighbor sites violates the necessary cubic symmetry around thorium~\cite{Dessovic2014, Pimon2022}, suggesting that the charge compensation for site~1 cannot be attributed to these positions.

We explored this setting using DFT simulations. By introducing a calcium vacancy or two fluorine interstitials at increasingly distant lattice sites, we observe that thorium displays a 4+ charge state while the band gap remains large, which is necessary for maintaining the transparency of the doped crystals in the visible and UV regimes.

\begin{figure}[htbp]
    \centering
    \includegraphics[width=8.0cm]{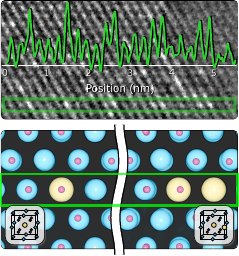}
    \caption{ {\bf Structural analysis of thorium doped calcium fluoride}. \textbf{Top}, STEM image and trace along the indicated line. CaF$_2$ seen from the [111] displays hexagonal symmetry as seen in the image. The scattered electron signal approximately doubles for a column containing thorium. Tentatively, higher-order cluster structures can be observed. Variation in signal intensity is due to misalignment, distortion, and charging of the material. See Methods for detailed experimental procedure and full field image of the sample (Fig.~\ref{fig:fullstem}). \textbf{Bottom}, Structural assignment of the two dominant microscopic sites, showing DFT optimized ionic positions from the [111] axis (inlay: sketch of the defect including the CaF$_2$ conventional cell). The green horizontal 
    line corresponds to the trace of \textbf{Top}. Left: Site 1. Right: Site 2.}
\label{fig:cluster}
\end{figure}

Furthermore, with increasing separation of thorium and the charge-compensating site, the local $O_h$ symmetry is restored, and the EFG approaches zero (see Methods for a quantitative study). Since other compensation schemes can be ruled out, we attribute the central zero EFG spectroscopy transition (site~1) to a charged thorium defect without local compensating atoms ($\ch{Th_{Ca}} - 2 \ch{e}$, see Fig.~\ref{fig:cluster}). Nalikowski et al.~\cite{Nalikowski2025}, using multiconfigurational theory simulations, attributed the same defect geometry to one of the features in Th:CaF$_2$ VUV absorption measurements~\cite{Beeks2024}.

The second dominant feature in the spectrum (site~2) is characterized by an asymmetry parameter of $\eta \approx 0.6$. Systems with more than threefold rotational symmetry would yield $\eta = 0$, ruling out compensation schemes featuring two interstitial fluorines, which exhibit $C_{3v}$ symmetry~\cite{Takatori2025}. Oxygen impurity schemes, such as \ch{2 O_F} or $\ch{O_F} + \ch{F_i}$, produce EFGs that are too small~\cite{Dessovic2014}, while \ch{O_i} has $C_{4v}$ symmetry. A nearest neighbor $\ch{v_{Ca}}$ with $C_{2v}$ symmetry has an incorrectly signed EFG and lower asymmetry, whereas a next nearest neighbor $\ch{v_{Ca}}$ has $C_{4v}$ symmetry. 
Consequently, no single thorium defect can explain the EFG extracted for site~2.
\vspace{5mm}

To investigate further, we performed scanning transmission electron microscopy (STEM) measurements using the high-angle annular dark-field (HAADF) technique on wedge shaped $\sim$20\,nm thickness samples (see Methods) of Th:CaF$_2$ (V14, concentration $2.6\cdot10^{20}$\,cm$^{-3}$, see\cite{Beeks2022}). The electron beam damage to the crystal is knock-on type, inducing some sputtering but not lattice disorder, which thus keeps the structure intact\cite{Jiang2012}. We imaged single columns of $\sim 20$ Ca atoms in the [111] crystallographic direction, seen in Fig.~\ref{fig:cluster}. Due to the larger nuclear charge of Th ($Z=90$) as opposed to Ca ($Z=20$), the Rutherford scattering of electrons is $90^2/20^2\sim 20$ times stronger, which allows us to clearly identify columns containing thorium. It is observed that columns containing Th generally cluster, indicating that it is energetically favorable to place Th as nearest neighbors in crystal growth. The data suggests that Th ions prefer being nearest neighbors at high concentrations, but atomic resolved STEM tomography is required to observe the 3-D nature, which is practically infeasible for highly radiation sensitive CaF$_2$. Clustering of dopants and distant charge compensation in CaF$_2$ is observed for all lanthanides\cite{Ma2020,Lacroix2014}.

Thus, we proceeded with a similar approach to the zero EFG case. We performed DFT simulations on thorium clusters by placing two thorium atoms at nearest neighbor Ca positions, removing four electrons from the simulation cell, and relaxing the system. The obtained EFG is 95\,V/\AA$^2$ and $\eta=0.6$, consistent with the experimental observations of this work and in refs.\cite{Zhang2024, Higgins2025}. Thus, we assign this defect structure to a cluster of two thorium atoms without local charge-compensating atoms ($\ch{Th_{Ca}} - \ch{4 e}$, see Fig. \ref{fig:cluster}).

Having accounted for the two most prominent sites~1 and 2, two more defect configurations remain to be identified. The scaling of site~3 and site~4 contribution with doping concentration (Table~\ref{tab:relative_Th_amount})  again indicates defects involving multiple thorium ions. For brevity, we leave this analysis for a future publication.

%%%%%%%%%%%%%%%%%%%%%%%%%%%%%%
\section{Discussion and Conclusion}
%%%%%%%%%%%%%%%%%%%%%%%%%%%%%%

The high-symmetry site~1, with the strongest signal, the most effective laser quenching and vanishing EFG, appears as a promising candidate for a solid-state nuclear clock. It was however not prominently observed in ref.~\cite{Zhang2024}, possibly indicating inhomogeneous broadenings different from site~2. In general, probing different nuclear quadrupole transitions in different defects may be used to eliminate systematics in clock interrogation sequences, i.e. in co-thermometry or to monitor stress. 

Laser induced quenching has been reported in ref.\cite{Schaden2025}, using a broad ($\le$10\,GHz) VUV laser, probably exciting \Th in all doping sites. The C10 crystal was found to be more susceptible to quenching than the higher concentration X2 crystal over a broad range of temperatures. We conjecture that this can be explained by a combination of the site-dependent quenching and the change of site occurrence with doping concentration observed in this work. A detailed characterization of LIQ for different microscopic doping sites will be the focus of future investigations. 

% Conclusion
In conclusion, we performed nuclear laser spectroscopy of \Th in three different CaF$_2$ crystals and identified four different microscopic sites, with distinct characteristic EFGs.  
In all crystals, two dominant sites contributed to more than 90\% of the obtained VUV signal. 
While the radiative lifetime appeared unaffected by different microscopic site configurations, the quenching efficiency varied strongly.  
By combining experimental data with DFT calculations, we assigned microscopic models to these two dominant sites. 
The laser Mössbauer spectroscopy introduced here can readily be transferred to other VUV transparent host materials such as single crystals or films, or opaque materials, when combined with conversion electron detection~\cite{Elwell2025}. 

%%%%%%%%%%%%%%%%%%%%%%%%%%%%%%%%%%%%%%%%%%%%% 

%TC:ignore

\section*{Acknowledgements}
This work was supported by JST, CREST Grant Number JPMJCR24I6, Japan, and supported by JSPS KAKENHI Grant Numbers JP21H04473, JP24K00646, JP24H00228, JP24KJ0168, JP25K17413, and JP25H00397. 
This work was also supported by JSPS Bilateral Joint Research Projects No. 120242002. 
This work has been funded by the European Research Council (ERC) under the European Union’s Horizon 2020 and Horizon Europe research and innovation programme (Grant Agreement No. 856415 and No. 101087184) and the Austrian Science Fund (FWF) [Grant DOI: 10.55776/F1004, 10.55776/J4834, 10.55776/ PIN9526523]. 
The project 23FUN03 HIOC [Grant DOI: 10.13039/100019599] has received funding from the European Partnership on Metrology, co-financed from the European Union’s Horizon Europe Research and Innovation Program and by the Participating States. 
K.B. acknowledges support from the Schweizerischer Nationalfonds (SNF), fund 514788 “Wavefunction engineering for controlled nuclear decays".
The computational results have been achieved in part using the Austrian Scientific Computing (ASC) infrastructure. 
We extend our gratitude to Stefaan Cottenier for fruitful discussions on symmetry properties of the electric field gradient and computational methods for calculating it.

\clearpage

\appendix

\section*{Appendix}
\setcounter{section}{0}
\renewcommand{\thesection}{\Alph{section}}

%%%%%%%%%%%%%
\section{Nuclear quadrupole splitting of \Th}
%%%%%%%%%%%%%
The Hamiltonian describing the interaction between the nuclear quadrupole moment $Q$ and the electric field gradient created by surrounding atoms can be written as 
\begin{equation}
H_{E2} = \frac{QV_{zz}}{4I(2I-1)} \left( 3I_z^2 - \bm{I^2} + \eta(I_x^2 - I_y^2) \right) 
\end{equation}
where $\bm{I} = (I_x,\ I_y,\ I_z)$ is the nuclear spin operator. 
If $\eta\neq0$, $H_{E2}$ is a non-diagonal matrix. 
The energy eigenvalues and eigenvectors of $H_{E2}$ are used for the calculation of the frequency splitting and transition strengths. 
Here, eigenstates of the isomeric state $\ket{M'}$ and the ground state $\ket{M}$ are written as 
\begin{align}
\ket{M'}=\sum_{m'}c_{M'm'}\ket{m'}, \ \  \ket{M}=\sum_{m}c_{Mm}\ket{m},
\end{align}
where $m$ ($m'$) is the magnetic quantum number for the nuclear spin of the ground (isomeric) state. 

For the isomeric state ($I_e=3/2$) energy eigenvalues are 
\begin{align}
  E'_Q(\ket{\pm3/2}) &= \frac{Q_eV_{zz}}{4}\sqrt{1+\frac{\eta^2}{3}}, \\
  E'_Q(\ket{\pm1/2}) &= -\frac{Q_eV_{zz}}{4}\sqrt{1+\frac{\eta^2}{3}}.
\end{align}
For the ground state ($I_g=5/2$) energy eigenvalues are the solutions of the following cubic equation:
\begin{align}
  x^3-28(3+\eta^2)x-160(1-\eta^2)=0,
\end{align}
where $E_Q=x \cdot Q_gV_{zz}/40$.
In the case of $V_{zz}>0$, $E_Q(\ket{\pm5/2})>E_Q(\ket{\pm3/2})>E_Q(\ket{\pm1/2})$ holds. 
The relative transition strength of a nuclear $M1$ transition from $\ket{M}$ to $\ket{M'}$ induced by a VUV laser is written as 
\begin{align}
  p(\ket{M}\to\ket{M'}) = \sum_{q=-1}^1|\braket{I_gm1q|I_em'}c_{M'm'}c_{Mm}|^2,
\end{align}
where $\braket{I_gm1q|I_em'}$ is the Clebsch-Gordan coefficient and $q=m'-m$.
Note that in the current experimental conditions where no magnetic field is not applied, $\ket{+M}$ and $\ket{-M}$ are degenerate and the relative transition strength between $\ket{\pm M}$ and $\ket{\pm M'}$ is the sum of those of degenerate states. 

%%%%%%%%%%%%%%%%
\section{Density Functional Theory Simulations}
%%%%%%%%%%%%%%%
We employed the Vienna Ab-initio Simulation Package (VASP)~\cite{vasp1, vasp2, vasp3, vasp4, vasp5} version 6.4.3 to perform density functional theory (DFT) simulations, leveraging the R2SCAN~\cite{r2scan} approximation to describe the exchange-correlation potential. To investigate the dilute doping limit, we constructed a $3 \times 3 \times 3$ supercell based on the conventional CaF$_2$ unit cell, yielding a cubic structure with an approximate edge length of \SI{16.3}{\angstrom} and comprising 324 atoms. The large cell size enables efficient simulation evaluation at the $\Gamma$-point only in reciprocal space. By optimizing lattice vectors separately from atomic positions, we ensured a more accurate representation of the defect geometry, and set convergence criteria as follows: We employed a threshold of \SI{0.008}{\electronvolt \per \angstrom} for ionic forces, while we halted optimization when the total cell energy difference between successive steps decreased to \SI{0.001}{\milli \electronvolt}. We used the pseudopotentials \texttt{F\_GW\_new}, \texttt{Ca\_sv\_GW}, and \texttt{Th} with an energy cutoff of 675 eV.

In Table~\ref{tab: distance}, we report the results of our investigation into distant charge compensations. By systematically introducing calcium vacancies at increasingly distant lattice sites, we observe that the EFG approaches zero, the band gap remains large, but the total energy increases. Given this analysis, we stress that we do not postulate an explicit mechanism of charge compensation; rather, we demonstrate that positioning the compensation at a sufficient distance is sufficient to maintain lattice symmetry. However, our simulations indicate that adjacent compensation mechanisms are lower in energy, which suggests that kinetic effects dominate during crystal growth. It is worth noting that CaF$_2$ exhibits a superionic phase near its melting point, where fluorine ions display high mobility while the calcium lattice remains rigid\cite{Beeks2024}.
\begin{table}[htbp] 
    \centering
    \caption{Increasing the distance $d$ of a calcium vacancy to the thorium site. $E$ is the internal electronic energy, referenced to a nearest neighbor calcium vacancy, $E_g^{\text{DFT}}$ is the DFT band gap, and $V_{zz}$ together with $\eta$ describe the EFG on thorium. For comparison, we also investigated the effect of two interstitial fluorines placed at distant sites, obtaining analogous results.}
    \sisetup{round-mode=places, round-precision=1}
    \vspace{10pt}
    \begin{tabular}{r r r r r}
        \hline
         $d$ (\r{A}) & $E$ (eV) & $E^{\text{DFT}}_g$ (eV) & $V_{zz}$ (\si{\volt \per \square \angstrom})  & $\eta$  \\
         \hline \hline
 \num{3.8} & \num{0.000} & \num{6.388} & \num{-109.804} & \num{0.083} \\
 \num{5.4} & \num{0.504} & \num{6.298} &   \num{50.493} & \num{0.000} \\
\num{12.7} & \num{1.072} & \num{5.789} &   \num{-0.673} & \num{0.010} \\
    \hline
    \end{tabular}
    \label{tab: distance}
\end{table}

%%%%%%%%%%%%%%%%
\section{Frequency scan}
%%%%%%%%%%%%%%%
By adjusting the frequency of the 786\,nm ECDL, the VUV frequency is scanned around the excitation frequency of \Th in 10\,MHz steps for each crystal. 
At each frequency point, the irradiation time is 60~seconds, and the subsequent measurement time is 300~seconds. 
For C10 and C13, a frequency range of 1.2\,GHz is scanned, while for X2 a range of 1.8\,GHz is scanned. 
The frequency range of X2 is wider to observe the weak transitions between $m = \pm 1/2\rightarrow m'=\pm 3/2$ from site~3 and site~4. 
The background rejection criterion is based on the peak height of the smoothed waveform from the PMT used for background rejection, acquired within $\pm$2\,$\mu$s of the trigger timing. 
After each spectra measurement, radioluminescence data are collected to estimate the background event rate ($p_2$) after background rejection. 
For each frequency point, the time spectrum, after background rejection without efficiency correction is fitted using the function $p_0\exp(-t/p_1)+p_2$, where $p_1$ = 630\,s and $p_2$ are fixed. 
The obtained $p_0$ value for each frequency is corrected for the residual signal effect according to:
\begin{align}
  c(f) = p_0(f) -  p_0(f+\Delta f) \exp\left(-\frac{t_{\mathrm{diff}}}{p_1}\right),
\end{align}
where $c(f)$ is the corrected signal intensity, $p_0(f+\Delta f)$ is the $p_0$ value from the previous frequency measurement, and $t_{\mathrm{diff}}$ is the time difference between the two consecutive measurements. 
Each $c(f)$ is normalized by the corresponding VUV intensity measured by the photodiode. 

The average EFG parameters calculated from the fitting are summarized in Table~\ref{tab:spectra_fit_Vzz}. 
The $V_{zz}$ value for site~1 is close to zero; consequently, all six transition lines overlap within the experimental uncertainty, rendering a determination of the asymmetry parameter $\eta$ impossible. 
The statistical uncertainties from the fits are smaller than the systematic uncertainties, likely dominanted by frequency fluctuations associated with the wavemeter and the fiber switch. 

The absolute value of the frequency deviates by at most about 100 MHz from the value reported in ref.\cite{Higgins2025}. 
This deviation appears to originate from the accuracy limitations of the wavemeter. 
Differences of the unsplit frequencies with reference to that for site~2 are within 20\,MHz, as summarized in Table.\ref{tab:spectra_fit_center}. 

For the X2 crystal with the highest $^{229}$Th concentration, the agreement between the data and the fitted spectrum is lower than for the other two crystals, which might indicate the presence of a fifth or further doping sites. 
However, the EFGs of the fifth site could not be determined precisely because its peaks overlap with those from other sites. 

\begin{table}[htbp] 
\caption{\textbf{${V_{zz}}$ (V/$\mathring{\mathrm{A}}$$^2$) and ${\eta}$ obtained by spectrum fitting}. Values in parentheses are differences of fitting results of the two spectra.} 
\vspace{10pt}
\centering
	\begin{tabular}{cccccc}  
		\hline
		target & EFG & site~1 & site~2 & site~3 & site~4\\
		\hline \hline
        C10 & $V_{zz}$ & 1(1) & 106(1) & $-319$(3) & 261(1) \\
         & $\eta$ & - & 0.59(1) & 0.22(16) & 0.00(0) \\ \hline
        C13 & $V_{zz}$ & 0(0) & 107(1) & $-319$(3) & 259(1) \\
         & $\eta$ & - & 0.61(3) & 0.05(9) & 0.00(0) \\ \hline
		X2 & $V_{zz}$ & 2(5) & 106(2) & $-320$(3) & 258(3) \\
         & $\eta$ & - & 0.60(1) & 0.18(5) & 0.21(6) \\
		\hline
	\end{tabular}
	\label{tab:spectra_fit_Vzz}
\end{table}

\begin{table}[htbp] 
\caption{ \textbf{Difference of the unsplit frequencies (MHz) with reference to that for site~2.} Values in parentheses are differences of fitting results of the two spectra.} 
\vspace{10pt}
\centering
	\begin{tabular}{cccc}  
		\hline
		target & site~1 & site~3 & site~4\\
		\hline \hline
        C10 & 3(1) & $-6$(2) & $-2$(1) \\
        C13 & 4(4) & $-4$(5) & $-3$(2) \\
		X2 & $-1$(1) & $-12$(8) & $-11$(3) \\
		\hline
	\end{tabular}
	\label{tab:spectra_fit_center}
\end{table}

%%%%%%%%%%%%%%%%
\section{Lifetime measurement without quenching} 
\begin{figure}[htbp]
\centering
\includegraphics[width=9.0cm]{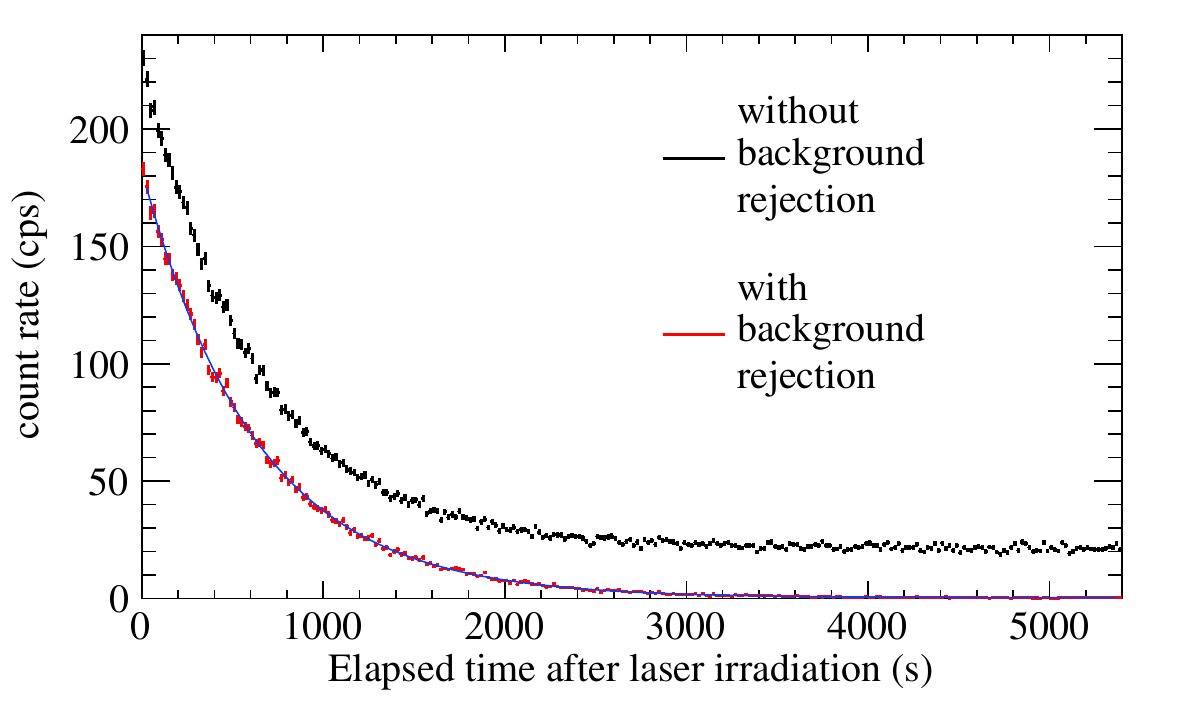}
\caption{{Example of a decay curve for purely radiative decay of \Thm.} 
Here, the efficiency correction is not applied and C13 is irradiated with the VUV frequency corresponding to the excitation frequency of site~1. 
The blue curve represents the fitted curve. 
The error bars represent the 68\% confidence interval of the statistical uncertainty. 
}
\label{fig:decay_curve}
\end{figure}

%%%%%%%%%%%%%%%
Figure~\ref{fig:decay_curve} shows an example of decay curves with and without radioluminescence background rejection. 
The amplified VUV PMT signal is combined with the clock generator signal and directed to the trigger channel of the oscilloscope for data acquisition. 
During the lifetime measurements, a 200\,Hz clock signal is used for monitoring the data acquisition efficiency and the background rejection efficiency. 
Lifetime measurements are conducted repeatedly under each condition.
The combined time spectra after background rejection with efficiency correction are fitted by $p_0\exp(-t/p_1)+p_2$, where $p_0$, $p_1$, and $p_2$ are the fitting parameters. 
The efficiency correction shifts the fitted lifetime typically by a few seconds.
Table~\ref{tab:lifetime_summary} summarizes the lifetime measurements of \Thm without irradiation of the quenching laser. 
For each condition, the lifetime is estimated to be approximately 630\,s, with no clear discrepancies observed.

\begin{table}[htbp] 
\caption{Lifetime measurement of \Thm without irradiation of the quenching laser. 
One measurement time for site~1 and 2 is 5400\,s and that for site~3 and 4 is\,3600 s. 
The errors represent the 68\% confidence interval of the statistical uncertainty. } 
\vspace{10pt}
\centering
	\begin{tabular}{cccc}  
		\hline
		site & target & set & lifetime (s)\\
		\hline \hline
		1 & X2 & 3  & $628.9\pm1.4$ \\
         & C13 & 3  & $626.9\pm1.3$ \\
         & C10 & 4  & $630.4\pm1.4$ \\ \hline
		2 & X2 & 8  & $629.9\pm1.1$ \\
         & C13 & 8  & $633.1\pm2.0$ \\
         & C10 & 9  & $632.6\pm2.6$ \\ \hline
		3 & X2 & 16 & $619.4\pm5.1$ \\
         & C13 & 16 & $600.2\pm20.0$ \\ \hline
        4 & X2 & 14 & $619.0\pm2.7$ \\
         & C13 & 7 & $632.0\pm9.1$ \\
		\hline
	\end{tabular}
	\label{tab:lifetime_summary}
\end{table}

\FloatBarrier
%%%%%%%%%%%%%%%%
\section{Lifetime measurement with quenching} 
%%%%%%%%%%%%%%%

Following VUV laser irradiation, the VUV signal rate is recorded for approximately 200\,s. 
Subsequently, the quenching laser is irradiated for approximately 200\,s. 
The rate measurement, interrupted during the quenching laser irradiation, is then resumed. 
To estimate the lifetime during the quenching ($\tau_q$), each decay curve after the background rejection analysis is fitted by $p_0\exp(-t/p_1)+p_2$, where $p_1$ is fixed at 630\,s. 
From the fit result, signal rates immediately before ($p_b$) and after ($p_a$) the quenching laser irradiation are determined. 
The fit region after the quenching laser irradiation starts approximately 180\,s after the end of the quenching laser irradiation to avoid effects from crystal heating. 
$\tau_q$ is calculated as 
\begin{equation}
    \tau_q = \frac{\Delta t}{\ln(p_b/p_a)},
\end{equation}
where $\Delta t$ is the duration of the quenching laser irradiation. 

%%%%%%%%%%%%%%%%
\section{Transmission electron microscopy}
%%%%%%%%%%%%%%%
To produce atomic resolved images of Th:CaF$_2$, we used the double corrected 60--300\,kV FEI Titan Themis transmission electron microscope at the EPFL CIME. We used high angle annular dark field imaging (HAADF) to image the atomic columns. To produce samples of Th:CaF$_2$ that were thin enough, we used tripod polishing\cite{Cha2016} on a crystal grown at the TU Wien to produce a wedge-shaped sample (Crystal V14, grown and imaged in the [111] direction\cite{Beeks2022}) that was $\sim$20\,nm thick close to the edge. The polished crystal was so thin, that the CaF$_2$ curled up at the edge. We centered the beam where the crystal was approximately 20\,nm thick, as any thinner target area would become amorphous and curl up. In the [111] axis, 20\,nm of material means approximately 20 atoms of Ca should be present in each column we image as the distance between 2 Ca ions in the [111] direction is \SI{9.56}{\angstrom}. Fluorine ions were not visible as they scatter too few electrons to be detectable. The scattering of electrons is proportional to the atomic number squared (Z$^2$), therefore F ions are $9^2/20^2\sim0.2$ times weaker and Th ions are $90^2/20^2\sim20$ times brighter than Ca ions. 

Before measurements, we cleaned the crystal using a gentle flow of oxygen plasma. Argon ion bombardment cleaning was attempted before but damaged the crystal and created dislocation loops, see\cite{Beeks2022}, most likely centered around the thorium lattice sites as these have been observed in the past\cite{Teodorescu1979}. Atomic resolution was not achieved for the ion bombarded sample, so we are uncertain if the dislocation loops center around the thorium. Considering that the number density of loops was of the order of the thorium concentration and it was observed in the literature\cite{Teodorescu1979}, it was concluded that the dislocation loops are at the thorium dopant site.

If the same location was imaged for a prolonged duration, the electron beam would sputter the sample. Therefore, we had limited time to put the sample into focus, on the right axis and achieve high resolution. We used a 20\,mrad collection angle, 300\,kV electrons, 5\,$\mu$s integration time and a 100\,pA current. By keeping the current and integration time as low as possible, we were able to image the single atomic columns.

\begin{figure}
    \centering
    \includegraphics[width=8.0cm]{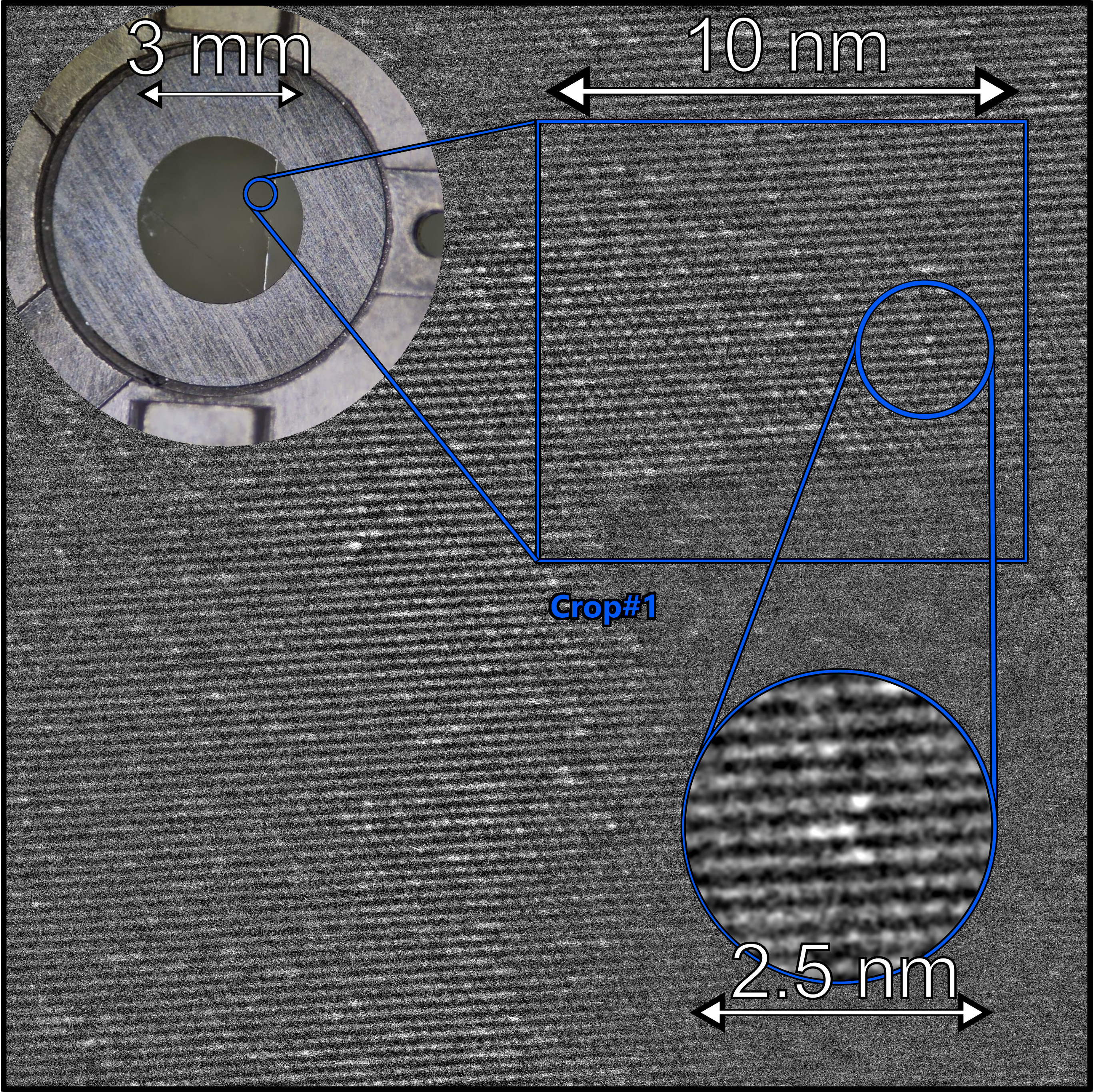}
    \caption{\textbf{Magnification steps and full field image of V14 with atomic resolution} Top left: Optical microscope image of the thinned single crystal in a TEM holder. Top right: Crop that was used to produce the background and trace of Fig.~\ref{fig:cluster}. Bottom right: Zoom of a 4 cluster of thorium. Background: Full field image where many different types of clusters of thorium can be discerned, a high-pass filter was applied using the proprietary Gatan software. Not all regions of the image show good contrast due to sputtering, deformations etc.}
    \label{fig:fullstem}
\end{figure}

To produce the images, we used the proprietary software of Gatan. We applied a Gaussian blur and high-pass filter. After which we cropped out a section of the total image, as not the entire field of view was in focus, see Fig.~\ref{fig:fullstem}. In this image, a large area that was thinner by sputtering can be observed (darker area, bottom right). Additionally, some areas are blurry and the individual columns could not be detected. Because the material was not perfectly straight, due to bending etc., not all areas could be in focus simultaneously.

%TC:endignore
\end{document}